\documentclass[journal]{IEEEtran}

\newtheorem{CC}{Corollary}
\newtheorem{PP}{Proposition}
\newtheorem{DD}{Definition}

\ifCLASSINFOpdf
\else
\fi

\ifCLASSOPTIONcompsoc
\usepackage[nocompress]{cite}
\else
\usepackage{cite}
\fi

\usepackage{amsmath}
\usepackage{amssymb}
\usepackage{graphicx}
\usepackage{epstopdf}
\usepackage{mathtools}
\usepackage{xcolor}

\begin{document}

\title{Prime and Co-prime Integer Matrices}


\author{
  Xiang-Gen Xia, \IEEEmembership{Fellow}, \IEEEmembership{IEEE},
  and Guangpu Guo

       \thanks{The authors are with the Department of Electrical and Computer Engineering, University of Delaware, Newark, DE 19716, USA (e-mail:
         \{xianggen, guangpu\}@udel.edu).
This work was supported in part by the National Science Foundation (NSF)  of USA under Grant CCF 2246917. }

}

\date{}

\maketitle


\begin{abstract}
  This paper investigates  prime and co-prime
  integer matrices and their properties.
  It characterizes all pairwise co-prime integer matrices that are also
  prime integer matrices.  
  This provides a simple way to construct families of pairwise co-prime
  integer matrices, that may have applications in multidimensional
  co-prime sensing and multidimensional Chinese remainder theorem. 
\end{abstract}

\begin{IEEEkeywords}
  \textit{Prime integer matrices, Gaussian integers, Gaussian primes,
    co-primality, Chinese remainder theorem (CRT), co-prime sensing.}
\end{IEEEkeywords}

\section{Introduction}\label{sec1}

It is well-known that prime integers are the most important subjects
in mathematics and have played vital roles in applications, such as
cryptography, digital communications, and signal processing etc. It plays a key role in the
Chinese remainder theorem (CRT)  that is to reconstruct a large integer
from its remainders modulo several small moduli and has many applications
in cryptography, digital communications and signal processing
as well \cite{crt1}. For a given set of moduli, the largest integer
that can be uniquely reconstructed from its remainders is smaller than 
the least common
multiple (lcm) of all the moduli and the lcm achieves the maximal
if all the moduli are pairwise co-prime, when the moduli sizes are upper
bounded. This means that it is critical to have a large set of
pairwise co-prime integers. 

The conventional prime integers have been generalized to other prime
algebraic integers in algebraic number fields with the corresponding CRT
\cite{Gauss1, congling1, congling2}. 
One example is Gaussian integers that are complex numbers with the conventional
integer real and imaginary parts. Prime Gaussian integers are called Gaussian primes
with simple and well-known necessary and sufficient conditions in terms of
the conventional prime integers (called rational primes) \cite{Gauss1}. In fact, an algebraic
integer can be equivalently represented by an integer matrix where all the elements  are the conventional integers. For a quadratic algebraic number field,
see, for example, \cite{congling1, congling2}, and in particular for a Gaussian integer of real and imaginary parts $a$ and $b$, respectively, its equivalent $2\times 2$ integer
matrix representation is $\left( \begin{array}{cc} a & -b \\ b & a\end{array}\right)$. Thus, a Gaussian prime also means that its equivalent
  $2\times 2$ integer matrix (representation) is prime. 
  Similarly, the CRT for Gaussian integers can be thought of as a CRT
  for $2\times 2$ integer matrices of the above special form, where Gaussian
  primes play the key role as well. It has recent applications
  in multi-channel self-reset analog-to-digital converter (SR-ADC) using
  modulo sampling for a complex valued bandlimited signal
  \cite{sradc, ccrt, SRADC}. 

  Prime integers have also been generalized to prime integer matrices
  in \cite{hua} where it obtains that an integer matrix is prime if and
  only if its determinant absolute value is a conventional prime integer.
  Recently, CRT has  been  generalized to general integer
  matrices of any form and any dimension in \cite{MD1, MD2} to reconstruct integer vectors from
  their vector remainders modulo several integer matrices called matrix moduli,
  i.e., multidimensional CRT (MD-CRT).
  Similar to the conventional CRT for integers, to have a large range
  of uniquely reconstructable integer vectors from their vector remainders,
  it is critical to have a large set of pairwise co-prime integer
  matrices as matrix moduli.
Also, co-prime
  integer matrices have applications in multidimensional
  sparse sensing \cite{coprime_pp}. 
  Unfortunately, 
  due to the non-commutativity of matrices, it is not obvious to construct
  a family of pairwise co-prime integer matrices of any dimension,
  although co-primality of integer matrices has been studied in
  \cite{pp_matrix3} mostly for dimensions $2$ and $3$.
    Interestingly,  a family of pairwise co-prime integer
  matrices was most recently
  obtained in \cite{guo1}
  for any dimension. Also, more detailed examples 
  on MD-CRT for separable and non-separable moduli can be found in \cite{guo1} as well. 

  It is known that any set of prime integers are pairwise co-prime.
  This holds for any prime algebraic  integers as well. Therefore, it is
  usually an easy way to construct families of pairwise co-prime elements
  by simply constructing families of
  prime elements, such as prime integers.
  This motivates us to use prime integer matrices to characterize and
  construct  pairwise co-prime integer matrices that are also prime. 

  In this paper, we first recall  prime integer matrices and 
  some of  their basic properties presented in \cite{hua}.
  We then investigate co-prime and pairwise co-prime integer
  matrices. 
  We  show that, similar to prime integers,
  two different prime integer matrices are co-prime.
  We then present the Hermite normal forms \cite{Hform}
  for prime integer matrices,
  and show that two prime integer matrices with the same determinant
  absolute value but different Hermite normal forms are co-prime. 
  Thus, it is
  easy to construct families of pairwise co-prime integer matrices
  by simply constructing families of prime integer matrices with
  different determinant absolute values that are prime and/or
  with the same determinant absolute value that is prime but different
  Hermite normal forms.
  This characterizes all pairwise co-prime integer matrices
  that are all prime. 
  It is interesting to note that none of the pairwise co-prime
  integer matrices in the construction in \cite{guo1} is prime when
  the dimension is above $1$. Also note that a different kind of primality
  for matrices was studied in \cite{pmatrix1, pmatrix2} for a given set of
  matrices. 

  Since Gaussian integers are special $2\times 2$ integer matrices,
  interestingly, 
  Gaussian primes and prime $2\times 2$  integer matrices are not exactly
  the same. When both real and imaginary parts of Gaussian integers
  are not $0$, Gaussian primes and prime $2\times 2$  integer matrices
  are the same. Otherwise, Gaussian primes may not be prime $2\times 2$
  integer matrices.
  However, two Gaussian integers are co-prime if and only if
  their equivalent $2\times 2$ integer matrices are co-prime. 
  It is well-known that any Gaussian integer
  has a unique Gaussian prime factorization \cite{Gauss1}. 
In contrast, an integer matrix can be uniquely factorized to a product of
  prime integer matrices when the order of the prime integer matrices
  in the product is not considered. Note that the order of primes in a
  product is not a problem if the primes are commutative, such as
  Gaussian primes (or their equivalent $2\times 2$ integer matrices).

  This paper is organized as follows. In Section \ref{sec2}, we
  study prime integer matrices and  their properties. We show
  their Hermite  normal forms and characterize all pairwise
  co-prime integer matrices that are all prime. 
  In Section \ref{sec3}, we present some connections between
  Gaussian primes and prime $2\times 2$ integer matrices.
  In Section \ref{sec4}, we conclude this paper.

\section{Definitions and Properties of Prime Integer Matrices}\label{sec2}

This paper considers integer matrices of size $D\times D$ for a positive
integer $D>1$, where all elements of matrices are integers.
Some notations are as follows.
$\mathbb{Z}$ denotes the set of all integers. 
All vectors and matrices
are $D$ dimensional integer vectors and $D\times D$ dimensional
integer matrices, respectively, 
unless
otherwise specified.
$I$ is the $D\times D$ identity matrix, and $0$
is also the all $0$ matrix or vector. $\det(A)$
denotes the determinant of matrix $A$.
And diag stands for a diagonal matrix. 

Below we introduce some necessary concepts
on integer matrices and for details, see, for example,
\cite{matrix,smith4}. 

\textbf{Unimodular matrix}: A square  matrix is called unimodular if its determinant is $1$ or $-1$. Otherwise, it is called non-unimodular. 

\textbf{Divisor and greatest common left divisor (gcld)}: A nonsingular integer matrix $A$ is a left divisor of a matrix $M$ if $A^{-1}M$ is an integer matrix. If $A$ is a left divisor of each of all $L \geq 2$  matrices $M_1, M_2, \dots, M_L$, it is called a common left divisor (cld) of
  $M_1, M_2, \dots, M_L$. Moreover, if any other cld is a left divisor of
  $A$, then $A$ is a {\em greatest common left divisor} (gcld) of
  $M_1, M_2, \dots, M_L$, denoted by gcld$(M_1, M_2, \dots, M_L)$. 

 \textbf{Co-prime matrices}: Two $D\times D$ matrices are left co-prime (or simply co-prime in this paper) if their gcld is a unimodular matrix. For two  matrices $M$ and $N$,  they are left co-prime if and only if
  the Smith form \cite{smith1, smith4} of the combined $D\times 2D$  matrix
  $(M \ N)$ is $(I \ 0)$. Also, it is not hard to see that if the determinant absolute values of two 
  matrices are co-prime, these two 
  matrices are co-prime.   
  Note that in this paper, only left coprimality is considered 
  and simply called coprimality. 

\textbf{Multiple and least common right multiple (lcrm)}: A  nonsingular
matrix $A$ is a right multiple of a matrix $M$, if there exists a nonsingular  matrix $P$ such that $A = MP$. If $A$ is a right multiple of each of all $L \geq 2$ matrices $M_1, M_2, \dots, M_L$, $A$ is called a common right multiple (crm) of $M_1, M_2, \dots, M_L$.
  Matrix $A$ is a {\em least common right multiple} (lcrm) of $M_1, M_2, \dots, M_L$, if any other crm of them 
  is a right multiple of $A$, denoted by lcrm$(M_1, M_2, \dots, M_L)$.
  From this definition, one can see that if $M$ is an lcrm of
  matrices  $M_1, M_2, \dots, M_L$, then $MU$ is also an lcrm of
  $M_1, M_2, \dots, M_L$ for any unimodular matrix $U$.
  It is not hard to see
  that if two matrices $A$ and $B$ are both lcrm of the same set of matrices
   $M_1, M_2, \dots, M_L$, then $A=BU$ for some unimodular matrix $U$. 
In addition,  for any groups of $D\times D$
  integer matrices $M_{1,1},\cdots,M_{1,L_1},
  \cdots, M_{k,1},\cdots,M_{k,L_k}$,
  we have
 \begin{equation}\label{lcrm1}
 \begin{aligned}
    &\text{lcrm}(M_{1,1},\cdots,M_{1,L_1},
    \cdots, M_{k,1},\cdots,M_{k,L_k}) \\
    = &\text{lcrm}( \text{lcrm}(M_{1,1},\cdots,M_{1,L_1}),
    \cdots,  \text{lcrm}(M_{k,1},\cdots,M_{k,L_k})).
 \end{aligned}
 \end{equation}
 
We next introduce prime matrices \cite{hua}, Chapter 14. 

\begin{DD} \cite{hua}
  A nonsingular matrix $A$ is called left prime (or simply prime) if
  it cannot be factorized to a product of two non-unimodular
  matrices. If $A$ is prime, then $A$ is called a prime matrix. 
\end{DD}

For the consistency, $1$ is considered as a conventional
prime integer, which corresponds to a unimodular
matrix treated as a prime matrix in this paper. 
It is clear that if $A$ is prime, then $AU$ and $UA$ are both
prime for any unimodular matrix $U$. Since the non-commutativity of matrices and
we only consider one side primality, i.e., the left primality. For
this reason, in this paper, a 
matrix $A$ and its right multiples $AU$ for unimodular matrices $U$
are counted the same (or associates \cite{hua})
for simplicity.
In other words, matrix $A$
and matrix $AU$ for any unimodular matrix $U$ are indistinguishable.
Clearly, matrices with different determinant absolute values are different.
In this sense, all the lcrm of a set of matrices  $M_1, M_2, \dots, M_L$
are the same, i.e., the  lcrm of a given set of matrices is  unique. 

\begin{PP}\label{P1}
  Two different prime  matrices are co-prime.
\end{PP}

    {\bf Proof}. Let $A$ and $B$ be two different prime matrices.
    If they are not co-prime, then their gcld matrix $C$ is not a unimodular
    matrix and $A=CA_1$ and $B=CB_1$ for two integer matrices $A_1$ and $B_1$.
    Since $A$ and $B$ are different, either $A_1$ or $B_1$ is not unimodular,
    which contradicts with the assumption that both $A$ and $B$ are prime.
    {\bf q.e.d.}

    \begin{PP}\label{P2} \cite{hua}
      A matrix is prime if and only if its determinant absolute
      value is a conventional prime integer.
    \end{PP}

    Its proof is not hard and can be found in \cite{hua}. 
%

        With Prop. \ref{P2}, we immediately have the following corollary.

        \begin{CC}\label{C1}
          If $A$ is a prime matrix, then its Smith form is
          $A=U\mbox{diag}(1,1,\cdots,1,\lambda_D)V$ where
          $U$ and $V$ are two unimodular matrices and $\lambda_D$ is a
          conventional prime integer.
        \end{CC}
        
        From Prop. \ref{P1}, we know that a set of different prime matrices
        are pairwise co-prime. Then, from Prop. \ref{P2},
        we have the following corollary.

        \begin{CC}\label{C2}
          Let $M_1, ..., M_L$ be $L$ different  matrices, i.e.,
          no two of them only differ by a unimodular matrix factor. If their
          determinant absolute values are conventional prime numbers,
          then they are pairwise co-prime.
          \end{CC}

        For   MD-CRT \cite{MD1, MD2} that have applications
        in multidimensional signal processing, it is important to
        have various families of pairwise co-prime integer matrices as mentioned
        earlier. 
        From Corollary \ref{C2},
 one can easily construct families of
 pairwise co-prime integer matrices by constructing
 matrices with different  determinant absolute values that are all prime
 integers. With this in mind, 
 one might want to ask what happens to the matrices with the same
 determinant absolute value that is a prime. In this case, any of them
 is prime but can they be pairwise co-prime?

 We next provide a complete
 answer to this question by using Hermite normal form of integer
 matrices (Chapter 14 of \cite{hua} and Chapter 14 of \cite{Hform})
 where it says that any matrix
 has a unique Hermite normal form. Combined with the primality of a
 matrix with the result in Corollary \ref{C1}, we have the following
 proposition.

 \begin{PP}\label{P7}
For a prime matrix $A$ that is not unimodular, its unique Hermite normal
form $H=(h_{mn})$ in $A=HU$ for some unimodular matrix
$U$ satisfies:
\begin{itemize}
\item[1)]
$H$ is a lower triangular matrix, i.e., $h_{mn}=0$ if $n>m$;
\item[2)] There is one and only one  diagonal element  that is not $1$, i.e.,
there exists one and only one $m_0$, $1\leq m_0\leq D$, such that
$h_{m_0m_0}=\lambda_D>1$ and $\lambda_D$ is a  prime integer,
and all $h_{mm}=1$ if $m\neq m_0$;
\item[3)] For this $m_0$ and any $m$th
  row of $H$ with $m\neq m_0$, all the elements of the $m$th row left to 
the one $h_{mm}=1$ on the diagonal are all $0$, i.e., $h_{mn}=0$
for any $1\leq n<m$;
\item[4)] All the elements $h_{m_0n}$ of the $m_0$th row left to 
   the diagonal $h_{m_0m_0}$ are nonnegative and strictly less than
  $h_{m_0m_0}=\lambda_D$, i.e.,
  $0\leq h_{m_0n}<h_{m_0m_0}$ for $1\leq n<m_0$.
\end{itemize}
The above Hermite normal form $H$ is shown as
\begin{equation}\label{2.1}
  H=\left( \begin{array}{ccccccc}
    1 & \cdots & 0 & 0         & 0 & \cdots  & 0 \\
      & \ddots &   & \vdots    & \vdots &         &\vdots \\
    0 &  \cdots& 1 & 0         & 0 & \cdots  & 0 \\
    a_1 & \cdots & a_k & \lambda_D & 0 &\cdots   & 0\\
    0 & \cdots & 0 &   0     & 1 & \cdots  & 0\\
    \vdots  &  &\vdots   & \vdots    &   & \ddots  & \\
    0 & \cdots & 0 &   0  & 0 & \cdots  & 1
  \end{array}\right),
\end{equation}
where $0<k<m_0$  and $0\leq a_1,...,a_k<\lambda_D$
are all integers and do not appear if $m_0=1$. 
 \end{PP}

 From the above Hermite normal form $H$ in (\ref{2.1}), one can see
 that there is only one row, i.e., the $m_0$th row $h_{m_0n}$,
 in $H$ that has one or  more 
 elements greater than $1$, and all the other rows have only one $1$
 in each row and all the other elements are $0$. 
One example of such a Hermite
normal form $H$ for $D=4$ is 
$$
\left( \begin{array}{cccc}
  1 & 0 & 0 & 0\\
  0 & 1 & 0 & 0\\
  2 & 0 & 3 & 0\\
  0 & 0 & 0 & 1
\end{array} \right).
$$

From the above Hermite normal form of a prime matrix, one difference
with the Smith form in Corollary \ref{C1} is that the element $h_{m_0m_0}=\lambda_D>1$ in the Hermite normal form does not have to be located
in the last row of the matrix as in the Smith form in Corollary \ref{C1}.
Since the Hermite normal form is unique for a matrix, we have the
following pairwise co-primality of the prime matrices with the same
determinant absolute values.

\begin{PP}\label{P8}
  All prime matrices that have the same determinant absolute value
  but different Hermite normal forms (\ref{2.1}) are different and
  pairwise co-prime.
\end{PP}

{\bf Proof}. Let $M_1$ and $M_2$ be two prime matrices and
$|\det(M_1)|=|\det(M_2)|$. If $M_1$ and $M_2$ are not co-prime,
then there exists a unimodular matrix $U$ such that $M_1=M_2U$.
This implies that $M_1$ and $M_2$ have the same Hermite normal form
and thus contradicts with the assumption.
{\bf q.e.d.}

As we can see from the Hermite normal form $H$ in (\ref{2.1}),
for any prime integer $\lambda_D>1$, its different
positions $m_0$ on the diagonal and different values of $a_1,...,a_k$
all produce different Hermite normal forms and therefore
different prime matrices, i.e.,  they are all pairwise
co-prime, from Props. \ref{P1} and \ref{P8}. 
Thus, 
from Corollary \ref{C2} and Prop. \ref{P8}, one is able to construct
many families of pairwise co-prime matrices with different
determinant absolute values that are prime and with
different Hermite normal forms but the same
determinant absolute value that is a prime.
The above results have also characterized all pairwise co-prime integer
matrices that are all prime. 
Interestingly, a family of pairwise co-prime integer matrices
of any dimension, none of which is prime when the dimension
is more than $1$, was obtained in \cite{guo1}. 

For the MD-CRT, similar to the conventional CRT, the range
of the uniquely determinable integer vectors from their
vector remainders depends on an lcrm of the matrix moduli.
Similar to the conventional integer case, we have the following
proposition.

\begin{PP}\label{P6}
If $L$ different matrices $M_1,M_2,...,M_L$
are prime and commutative,
i.e., $M_lM_m=M_mM_l$, $1\leq l,m\leq L$, 
  then their product is their lcrm, i.e.,
  $M_1 M_2\cdots M_L=\mbox{lcrm}(M_1,M_2,...,M_L)$.
\end{PP}

{\bf Proof}.
From Prop. \ref{P1}, we know that matrices $M_1,M_2,...,M_L$
are pairwise co-prime. For any two co-prime matrices $M_l$ and $M_m$ with 
$l\neq m$,
since they are commutative, from \cite{PP4}, i.e., the proposition 3 in
\cite{MD1}, we know that
$M_lM_m$ is their lcrm, i.e., lcrm$(M_l,M_m)=M_lM_m$.
Then Prop. \ref{P6}
is proved by the lemma 2 in \cite{MD1} and (\ref{lcrm1}). 
{\bf q.e.d.}

Although
it is not easy to study commutative matrices of a high dimension
in general, 
it is not hard to see that for $2\times 2$ matrices,
commutative matrices can be
always represented by
\begin{equation}\label{2.2}
 A(a,b) =  \left( \begin{array}{cc}
    a & -\alpha b\\
    b & a-\beta b \end{array}\right)
\end{equation}
in case $b\neq 0$, where $\alpha$ and $\beta$ are two fixed constants.
It is easy to check that two matrices $A(a_i,b_i)$, $i=1,2$, 
in (\ref{2.2}) are commutative.
Thus, for any set $\mathbb{S}$ of matrices $A(a,b)$ with either different
determinant absolute values that are prime or
the same determinant absolute value that is prime
but different Hermite normal forms are both pairwise
co-prime and commutative. Therefore, by Prop. \ref{P6} 
their lcrm is their product that determines the range
of the uniquely reconstructable integer vectors of dimension $2$ from
their vector remainders modulo the $2\times 2$ matrices
in $\mathbb{S}$ \cite{MD1}.

Interestingly, algebraic integers in a quadratic algebraic number field
with a minimal polynomial $r_2 x^2+r_1 x+r_0$ can be equivalently
represented by $2\times 2$ matrices of the form in (\ref{2.2})
with $\alpha=r_0/r_2$ and $\beta=r_1/r_2$, where $b$ can be $0$,
\cite{congling1, congling2}. Since algebraic integers are all complex numbers
and therefore naturally commutative. When $r_2=1, r_1=0, r_0=1$, the
above algebraic integers  become Gaussian integers that will be
especially studied in next section. 



        \section{Connection with Gaussian Integers and Gaussian Primes}\label{sec3}
        In this section, we investigate and
        provide some connections between $2\times 2$ integer matrices
         with Gaussian integers.
        
        Gaussian integers are represented by $a+jb$ where $a$ and $b$ are
        conventional integers in $\mathbb{Z}$ and $j=\sqrt{-1}$.
        The ring of Gaussian integers is denoted by $\mathbb{Z}[j]$.
        A Gaussian integer $z=a+jb$ can be equivalently represented
        by the following $2\times 2$
        integer matrix
        \begin{equation}\label{3.1}
          z=\left(\begin{array}{cc} a & -b\\ b & a\end{array}\right),
        \end{equation}
        which is called the matrix representation of Gaussian integer $z$. 

        A Gaussian integer is a {\em unit} if its absolute value is $1$. It is
        then not hard to see that a Gaussian integer is a unit if and only if 
        its matrix representation (or simply matrix) in (\ref{3.1})
        is a unimodular matrix.
        A Gaussian integer is prime if it cannot be factored to a product of 
        two non-unit Gaussian integers.
        When a Gaussian integer is prime, we call it a Gaussian prime.
        In the following, interestingly we will see that
        this primality of Gaussian integers is different from that of
        $2\times 2$ integer matrices defined in Section \ref{sec2} in general. 

        First it is well-known \cite{Gauss1}
        that, a Gaussian integer $a+jb$ with
        $ab\neq 0$ is prime if and only if its norm, i.e., $a^2+b^2$,
        is a conventional prime integer. Since $a^2+b^2$ is also the determinant absolute value of its matrix representation in (\ref{3.1}),
        from Prop. \ref{P2} this primality
        is the same as that of integer matrices in Section \ref{sec2}.
        However, a Gaussian integer $z=a+jb$ with $ab=0$ is prime if and only if
        $|a|$ or $|b|$ is a conventional prime integer that is equal to $3$
        modulo $4$. On the other hand, for a Gaussian integer $z=a+jb$ with
        $ab=0$, its matrix representation is either 
        $$
        \mbox{diag}(a,a) =\left( \begin{array}{cc} a & 0 \\0 & a\end{array} \right)
          \mbox{ or  }\left( \begin{array}{cc} 0 & -b\\b & 0 \end{array} \right)
          $$
          that is not prime if $z$ is not a unit. 
        For example, $3$ is a Gaussian prime, while
        its matrix representation
        $\mbox{diag}(3,3)$ is not prime. From the above analysis,
        we  have the following proposition.

        \begin{PP}\label{P3}
          For a Gaussian integer $z$, if its matrix representation is
          prime, then it is a Gaussian prime, and the opposite may not be true. 
          \end{PP}

        In other words, matrix primality implies Gaussian primality, and
        Gaussian primality may not imply matrix primality.
        One reason might be as follows. Since in terms of the matrix primality,
        a matrix can be equivalently converted to a diagonal
        matrix of the form
        in Corollary \ref{C1}
        by using the Smith form decomposition that may not always preserve
        the matrix representation structure (\ref{3.1}) for a Gaussian
        integer in the decomposition steps. This means that
        the Smith form in Corollary \ref{C1} for a prime matrix
        may not apply to a Gaussian prime, i.e., a Gaussian prime
        does not have to have the Smith form in Corollary \ref{C1}. 
        
        We next want to see the co-primality of two matrices and two
        Gaussian integers.
        First two Gaussian integers are called co-prime if they do not have
        any non-unit common factor. Although the primalities of
        a Gaussian integer and its matrix representation are not equivalent,
         the co-primalities of two Gaussian
         integers and their matrix representations are equivalent
         \cite{congling1}, which can be seen 
         by using Bezout's identities for both Gaussian integers and
         integer matrices. 

        \begin{PP}\label{P4} \cite{congling1}
          Two Gaussian integers are co-prime if and only if their
          matrix representations are co-prime in the integer matrix
          sense.
        \end{PP}

        From the above results, 
        it is not hard to see that two  Gaussian primes with different
        norms are co-prime, similar to the conventional prime
        integers. For two Gaussian primes with the same
        norm, their equivalent matrix representations have the same
        determinant absolute value. In this case, from the $2\times 2$ integer
        matrix point of view, two different prime integer matrices with the same
        determinant absolute value may be co-prime if their Hermite normal
        forms are different from Prop. \ref{P8}. If they only differ by a unit,
        then they are the same and thus not co-prime, and their
        equivalent matrix representations have the same Hermite normal
        form.
        If they do not differ by a unit, 
         they have to be co-prime. 
        Then, their equivalent matrix representations
        have different Hermite normal forms. An exmaple is 
        $4+5j$ and $4-5j$ that have the same norm but different Hermite
        normal forms and thus co-prime, and do not differ by a unit. 


        It is well-known \cite{Gauss1} that  any Gaussian integer
        has a unique prime factorization, i.e., any Gaussian
        integer can be uniquely factorized to a product of some prime
        factors. This is also true for
        matrix prime factorization if the order of the prime factors in the
        product is not considered.

        \begin{PP}\label{P5}
          A non-singular  matrix $A$ can be uniquely factorized to a product of
          prime matrices called prime factors, if the order 
           the prime factors in the product is not considered. 
        \end{PP}

        {\bf Proof}. Let $A=IP_1P_2\cdots P_L$ be a prime factorization of matrix $A$, where all $P_i$ are prime. For each $P_i$, let $P_i=U_i\Lambda_iV_i$ be
        its Smith form decomposition with unimodular matrices
        $U_i, V_i$, where $\Lambda_i$ is  a diagonal integer matrix, 
        $i=1,2,...,L$. Since all $P_i$ are prime, from Corollary \ref{C1}, 
        $\Lambda_i=\mbox{diag}(1,1,..., 1,\lambda_i)$ for some
        conventional prime integers $\lambda_i$. Since $|\det(A)|=\prod_i^L \lambda_i$ whose factorization
        is unique, and $IW_1,\Lambda_1W_2, ..., \Lambda_L W_{L+1}$ are the
        same as $I, \Lambda_1,...,\Lambda_L$ for all unimodular matrices
        $W_1,W_2,...,W_{L+1}$, respectively,
        the above factorization of matrix $A$
        is unique if the order of the factors in the product is not considered.
        {\bf q.e.d.}
        
 A similar integer matrix factorization can be found in \cite{hua}.         
Note that although the above connection
        is only studied for Gaussian integers, 
        similar studies may apply to other algebraic
        integers in other algebraic number fields.

        \section{Conclusion}\label{sec4}
        In this paper, we have applied prime integer matrices to
        construct and characterize all pairwise
        co-prime integer matrices that are all 
        prime. For instance, 
        two  prime integer matrices with different
        determinant absolute values that are prime are co-prime, and 
        two prime integer matrices with the same determinant absolute
        value that is prime but different Hermite normal forms are
        co-prime. 
        These pairwise 
        co-prime integer matrices  may have applications
        in multidimensional CRT and multidimensional sparse sensing.
        We have also investigated  the connections between
        prime integer matrices and Gaussian primes.


\end{document}